\documentclass[a4paper,11pt]{article}
\usepackage{pos}

\title{The Hubble constant tension: current status and future perspectives through new cosmological probes}

\author*[a,b,c]{Maria Dainotti}
\author[d,e]{Biagio De Simone}
\author[f,g]{Giovanni Montani}
\author[h,i]{Tiziano Schiavone}
\author[d,e]{Gaetano Lambiase}

\affiliation[a]{National Astronomical Observatory of Japan,\\
  2 Chome-21-1 Osawa, Mitaka, Tokyo 181-8588, Japan}

\affiliation[b]{The Graduate University for Advanced Studies (SOKENDAI), \\
2-21-1 Osawa, Mitaka, Tokyo 181-8588, Japan}

\affiliation[c]{Space Science Institute,\\ 
Boulder, Colorado}

\affiliation[d]{Department of Physics “E.R. Caianiello", University of Salerno,\\
Via Giovanni Paolo II, 132, Fisciano, I-84084 Salerno, Italy}

\affiliation[e]{INFN Gruppo Collegato di Salerno - Sezione di Napoli - c/o Dipartimento di Fisica “E.R. Caianiello”, Ed. F, University of Salerno \\
 Via Giovanni Paolo II, 132, Fisciano, I-84084 Salerno, Italy}

\affiliation[f]{ENEA, Fusion and Nuclear Safety Department, C.R. Frascati, \\
 Via E. Fermi 45, Frascati, I-00044 Rome, Italy}

\affiliation[g]{Physics Department, “Sapienza" Universilatexmk -pdf skeletonty of Rome, \\
 P.le Aldo Moro 5, I-00185 Rome, Italy}

\affiliation[h]{Department of Physics “E. Fermi", University of Pisa, \\
Polo Fibonacci, Largo B. Pontecorvo 3, I-56127 Pisa, Italy}

\affiliation[i]{INFN, Istituto Nazionale di Fisica Nucleare, Sezione di Pisa, \\
Polo Fibonacci, Largo B. Pontecorvo 3, I-56127 Pisa, Italy}

\emailAdd{maria.dainotti@nao.ac.jp}

\abstract{The Hubble constant ($H_0$) tension is one of the major open problems in modern cosmology. This tension is the discrepancy, ranging from 4 to 6 $\sigma$, between the $H_0$ value estimated locally with the combination of Supernovae Ia (SNe Ia) + Cepheids and the cosmological $H_0$ obtained through the study of the Cosmic Microwave Background (CMB) radiation. The approaches adopted in \cite{2021ApJ...912..150D} and \cite{Dainotti2022SNe} are introduced. Through a binning division of the Pantheon sample of SNe Ia \cite{2018ApJ...859..101S}, the value of $H_0$ has been estimated in each of the redshift-ordered bins and fitted with a function lowering with the redshift. The results show a decreasing trend of $H_0$ with redshift. If this is not due to astrophysical biases or residual redshift evolution of the SNe Ia parameters, it can be explained in light of modified gravity theories, e.g., the $f(R)$ scenarios. We also briefly describe the possible impact of high-$z$ probes on the Hubble constant tension, such as Gamma-ray bursts (GRBs) and Quasars (QSOs), reported in \cite{Dainotti2022SNe} and \cite{LenartBargiacchi2022}, respectively.}

\FullConference{%
  Corfu Summer Institute 2022 "School and Workshops on Elementary Particle Physics and Gravity",\\
  28 August - 1 October, 2022\\
  Corfu, Greece}

\begin{document}
\maketitle

\section{Introduction}
One of the major goals of modern science is to understand the structure of the universe, relying on the most accredited paradigm for describing its current state and future evolution. To date, the strongest candidate for such a purpose is the so-called $\Lambda$CDM model for cosmology, based on the standard FLRW metric \cite{1933ASSB...53...51L}. The basic hypothesis under this formulation are: the presence of the cold - namely, non-relativistic dark matter (CDM) - whose density, in combination with the baryonic matter, is described by the $\Omega_{0m}$ parameter; the acceleration of the universe driven by the dark energy which, in turn, is represented by the $\Omega_{0\Lambda}$ parameter. One of the greatest achievements by the $\Lambda$CDM is the capability of predicting the accelerated expansion phase of the universe, as proven by the works of Riess \cite{1998AJ....116.1009R} and Perlmutter \cite{1999ApJ...517..565P} through their analysis of SNe Ia, which are among the best standard candles so far discovered.\\
Despite all the great success that the $\Lambda$CDM achieved in the scientific community, there are still many unsolved problems related to this model. First of all, the nature of dark matter and dark energy is not clear. Second, the void energy measurement through quantum mechanics and the void energy estimated with the cosmological observations differ by 120 orders of magnitude (the so-called \textit{fine-tuning problem}). Third, in principle, there is no clear reason why the universe's dark matter and dark energy contributions must have the same magnitude order today - \textit{the coincidence problem} \cite{1989RvMP...61....1W,2003RvMP...75..559P}. It is worth mentioning also the \textit{lithium problem} \cite{FranchinoVinas2021}, namely the discrepancy between the predicted abundance of primordial lithium as inferred from the CMB observations and the local observations in the halo stars of our Galaxy. Finally, one of the most intriguing problems that affect the $\Lambda$CDM model is the so-called \textit{Hubble constant} ($H_0$) \textit{tension}: this is the discrepancy, in more than 4 $\sigma$, between the value of the Hubble constant estimated through the local probes \cite{riess2022comprehensive} and the cosmological value inferred with the power spectrum of the CMB \cite{2020A&A...641A...6P}. The solution to this tension is not an easy task, and many suggestions have been provided by the scientific community. Considering the theoretical scenarios, many researchers have proposed a modification of gravity  \cite{doi:10.1142/S0219887807001928,2021PDU....3200807L,2021arXiv210603093A,Gurzadyan2021,Asghari2021,Sivaram2021,Stepanian2022,Gangopadhyay2022,Sharma2022modif,fierz1956physical,2011PhR...509..167C,Ray2021,2018PhRvD..98d4023B,2019arXiv190401016A,2000PhRvL..85.4438A}, in particular the $f(R)$ theories \cite{2007PhRvD..75h3504A,2007PhRvD..76f4004H,2021NuPhB.96615377O,2007PhRvD..75d4004S,2006CQGra..23.5117S,2010RvMP...82..451S,2007JETPL..86..157S,2008PhRvD..77b3507T,2021arXiv210400596S,2018PhLB..777..286L,Nunes_2017,Sotiriou,Farrugia2021,VanKy2022,Park2022}, the $f(T)$ theories \cite{2021MNRAS.500.1795B,2020CQGra..37p5002E,2018JCAP...05..052N,2021arXiv210414065N,Nunes_2018,10.1093/mnras/staa3368,Ren2021,Najera2021,Najera2021fitting,Ren2022,Aljaf2022,Briffa2023,Santos2022,Mandal2022,Yang2022fT}, or the $f(Q)$ theories \cite{Koussour2022,Koussour2022fQ}.\\ It is also important to mention that many alternative cosmological models have been suggested by the scientific community \cite{1947MNRAS.107..410B,1974MNRAS.167...55B,2000GReGr..32..105B,2001IJMPD..10..213C,2019CQGra..36d5007C,2006JCAP...08..011G,2020arXiv201102858K,2019PhRvD.100j3524R,2001PhRvD..64f3501Z,2021EPJC...81..295H,2021arXiv210509790K,2019PhRvD.100b3532A,2021IJMPA..3650044P,1993PhRvD..48.3436D,2000PhRvL..85.2236B,2001PhRvD..63f3504E,2018LRR....21....1B,Li_2015,Shrivastava2021,Palle2021,Petronikolou2021,Aghababaei2021,Bansal2021,Parnovsky2021b,Normann2021,Singh2021,Pereira2021,Ben_Dayan_2013,Fanizza_2020,Buchert_2000,Gasperini_2009,Gasperini_2011,Gariazzo2021,Ambjorn2021,Banik2021,Lulli2021,Li2021spatial,Nilsson2021,cea2022,koksbang2019homogeneous,koksbang2019backreaction,DiValentinomachina,CyrRacine2021symmetry,HernandezAlmada2022entropy,Asvesta2022,Leonhardt2022,Colaco2022,Adhikari2022,Rasouli2022,Bhardwaj2022,Camarena2022nonCopernican,Sharma2022oscill,HernandezJimenez2022,Bruno2022,Banerjee2022,Kuzmichev2022,Seitz2022,Kaneta2022}.\\
The modifications can also be theorized on the fundamental constants, or physical observables \cite{1999PhRvD..59d3515B,2020arXiv201010292N,2021arXiv210109862L,Syksy2009,Syksy2010,Hart2021,Hart2021,Spallicci2022,Trivedi2022,Kalbouneh2023,Zhang2022}, in particular, the gravitational constant \cite{2020PhRvD.102b3529B,2005JCAP...05..003M,2020EPJC...80..570W,Ballardini2021,sakr2021,alestas2022,Marra2021,Perivolaropoulos2022dinosaurs,Maeda2022CuscutaGalileon,Perivolaropoulos2022symmetron}.\\
Of particular interest are the theoretical frameworks that consider the contribution of dark energy \cite{	2017PhRvD..96b3523D,2020JCAP...07..045D,2016PhRvD..94j3523K,2019EPJC...79..141L,2019ApJ...883L...3L,2020ApJ...902...58L,2018JCAP...09..025M,2019PhRvL.122v1301P,2019PhRvD..99d3543Y,2012IJMPD..2130002Y,2021PhRvD.103d3518T,2021arXiv210303815Y,2019EPJC...79..762S,2019IJMPD..2850152S,2020EPJC...80..826A,2020RAA....20..151G,2020MPLA...3550107G,2021PhLB..81235990H,2004PhRvD..70f3519C,Geng2021,Benisty2021,Joseph2021,Colgain2021DDE,hernandezalmada2021kaniadakis,Belgacem2021,Burgess2021,Prat2021,Karwal2021,Nojiri2021,firouzjahi2022cosmological,Beltran2021,Artymowski2021,Zhou2021,Allali2021,DiValentino2021minimaldarkenergy,MorenoPulido2021,perivolaropoulos2014large,perivolaropoulos2011Tolman,Banihashemi2021,Divalentino2018,Chang2022,Poulin2021,Sabla2022,delaMacorra2022,DiGennaro2022,Yao2022,Kojima2022,Bhardwaj2022hol,Yao2022ide,GomezValent2022ede,Rezazadeh2022,Simon2022,Trodden2022,McDonough2022,Cardona2022,Herold2022,Deledicque2022,Staicova2022,Brissenden2022,Lin2023,Ong2022,DiValentino2023,Shafieloo2022,Kamionkowski2022,Zhou2022,Murai2022}, the phantom dark energy \cite{2020PhRvD.101l3516A,2021arXiv210507103L,Bag2021,DiValentinodark}, the dark matter \cite{	Blinov2021,Sola2021,Hansen2021,gutierrezluna2021scalar,Jodlowski2020,Ghose2021,fernandez2021,Drees2021,safari2022,Okada2022,Anchordoqui2022,Archidiacono2022dm,Holm2022,BuenAbad2022,2016PhRvD..94l3525B}, the dark radiation \cite{Ghosh2021,Berghaus2022,Schoneberg2022}, the dark energy gravity models \cite{2016JCAP...05..067B}, the dark fluid models \cite{2020IJMPD..2950097A,2019MNRAS.490.2071Y}, and the models the dark energy and dark matter interaction between themselves or with the other sectors of the universe \cite{	2019JCAP...06..003A,2009PhRvD..79f3518C,2020IJMPD..2950057M,Hoshiya2022,Harko2022,Aich2022,Wang2022dmde,Mekuria2022,Zhang2022ide}. The running or interacting vacuum models are also a proposal for relieving the Hubble tension \cite{MorenoPulido2022,Peracaula2020,Peracaula2019,GomezValent2018,GomezValent2017,Sol2017,SolPeracaula2021,SolPeracaula2022a,MorenoPulido2020,MorenoPulido2022a,SolPeracaula2022,MorenoPulido2022,Kaeonikhom2022}.\\
Other ideas stem from the early time physics or inflation scenario modifications \cite{2020PhRvD.102j3525K,2019PhRvD..99d3514L,1983PhLB..129..177L,2002hep.ph...10162R,Ye2021a,Aloni2021,Rashkovetskyi2021,Khosravi2021,Leon2022inflation,Aboubrahim2022,Ildes2022,Huang2022,Matsumoto2022,Gaztanaga2022,Ye2022inflation,Lin2022,Bourakadi2022,Smith2022,Lee2022recomb}, late time modifications \cite{2021MNRAS.504.3956A,2021arXiv210514332E,Alestas2021c,Aghaei2021,Perivolaropoulos2021Lagrangians,Perivolaropoulos2021phantom,Divalentino2019,Heisenberg2022a,Heisenberg2022b,Cai2022}, fluctuations and perturbation theories \cite{2017PTEP.2017h3E04T,Belgacem2022,	2017PTEP.2017e3E01T,2018PTEP.2018b1E01T,2019arXiv190609519T,2020PTEP.2020a9202T,Tomonaga2022}, or even from the possibility of inhomogeneities in the universe, local underdensities or voids \cite{2014PhRvL.112v1301B,2020A&A...633A..19B,2019JCAP...09..006C,2019ApJ...875..145K,2020MNRAS.499.2845H,2013ApJ...775...62K,2020MNRAS.491.2075L,2019MNRAS.490.4715S,1934PNAS...20..169T,Cai:2020tpy,Martin2021,Castello2021,Wong2022,koksbang2020observations,koksbang2016,DiValentinophase,Hosking2022,Yusofi2022,Gurzadyan2022void}.\\
Other discussions have been provided taking into account a modification to the physics of standard model particles, like the neutrino, \cite{2021arXiv210600025D,GonzalezLopez2021,Duan2021,das2021selfinteracting,DiValentino2021neutrino,Naidoo2021,Khalifeh2021,Corona2021,Niedermann2021,DiValentinoneutrino,DiValentinosolution,Archidiacono2022,Sharma2022,GarciaArroyo2022}, the gravitino and gravitone conjectures \cite{2021arXiv210410181C,Piran2022} or the most exotic scenarios that involve strings \cite{1994NuPhB.423..532D}, axions \cite{Fung2021,Cuesta2021,Gu2021,Mawas2021,luu2021axihiggs,Ye2021resolving,Rudelius2022}, B-meson decays \cite{Ghosh2017}, lepton flavor universality violation \cite{Heeck2022}, and the Planck mass \cite{Benevento2022}.\\
Not only the theoretical scenarios have been called into question, but of course, the statistical analysis on the $H_0$ measurements constitutes a significant fraction of the literature regarding the Hubble tension \cite{2013PhRvL.110x1305M,2020Ap&SS.365..131M,Mercier2021,Divalentino2020,Wagner2022,LopezCorredoira2022,Gueguen2022,Jia2022,2020JCAP...12..018G,2021MNRAS.501.3421K,2020PhRvD.102b3520K}.\\ 
The most tricky discussion concerning the Hubble tension must be done considering the probes used for its measurement: Cepheids \cite{	2019ApJ...876...85R,2019ApJ...886L..27R,2019MNRAS.484L..64S,2021ApJ...911...12J,riess2022comprehensive,Mortsell2021,Mortsell2021hubble,romaniello2021,Efstathiou2022SH0ES,Riess2018reply,Breuval2022,Thakur2022,Yuan2022}, SNe Ia \cite{2014A&A...568A..22B,2013ApJ...770..107C,2011A&A...529L...4C,2010A&A...523A...7G,2010ApJ...715..743K,2017ApJ...836...56K,2010ApJ...722..566L,2019MNRAS.486.2184M,2016ApJ...822L..35S,2010MNRAS.406..782S,1998A&A...331..815T,2017MNRAS.471.2254Z,2021arXiv210514514T,2018EPJC...78..755D,2019MNRAS.486.5679Z,2021ApJ...912L..26B,2019ChPhC..43l5102C,2020ApJ...892L..28H,2020MNRAS.494.2158R,2021MNRAS.504.5164C,Krishnan2021b,Perivolaropoulos2021,Peterson2021,Rodney2016,Conley2011,Horstmann2021,scolnic2021pantheon,brout2021pantheon,carr2021pantheon,popovic2021pantheon,GomezValent2022,Brownsberger2021pantheon,Deibel2021SNe,Benisty2022SNe,Ballardini2022,Brout2022params,deVicenteAlbendea2022,Mazo2022,Kenworthy2022,Garnavich2022,GallegoCano2022,Tiwari2022,Wojtak2022,Lu2022sncal,Zhai2022sne,MullerBravo2022,Galbany2022,Dhawan2022bayesn,Cowell2023,Perivolaropoulos2023,Wang2022pantheonplus,DhawanMortsell2023}, lensed objects \cite{2019ApJ...887..163M,Qi2022lensedSNe,Shah2022SNe,	2016MNRAS.461.4099B,2020MNRAS.498.1420W,2019MNRAS.490.1913W,2021MNRAS.501.1823B,Liu2021b,Hou2021,Shah2022,Zhu2022}, Cosmic Chronometers (CCs) \cite{2012JCAP...08..006M,2016JCAP...08..011N,2010JCAP...02..008S,Borghi2021}, CMB \cite{	2003ApJS..148....1B,2013ApJS..208...19H,2016A&A...594A..13P,2020A&A...641A...6P,2020JCAP...09..055K,2020EPJC...80..369Q,2021arXiv210510425V,2021arXiv210503003T,Liu2021a,Luongo2021,DiValentinoplanck,Hazra2022,Takahashi2022,Ye2022,Hart2022,Hayashi2022,Jiang2022}, Cosmic Dawn observations \cite{Sarkar2022}, galaxies clusters and sky-surveys \cite{	2017MNRAS.470.2617A,2021MNRAS.500.5249A,2018MNRAS.476..151E,2015MNRAS.449..835R,2018PhRvD..98d3528T,2020ApJ...901...90A,RuizZapatero2021,Euclid,Mantz2021,Ghosh2022ska,Bian2023,Brieden2022}, large scale structures \cite{Zhang2021,fanizza2021precision,Gurzadyan2022}, galaxies parallax \cite{Ferree2021}, Baryon Acoustic Oscillations (BAOs) \cite{	2015PhRvD..92l3516A,2011MNRAS.416.3017B,Sharov_2018,2005ApJ...633..560E,2019JCAP...10..044C,Staicova2021,Kumar2022,Schoneberg2022baobbn}, Big Bang Nucleosynthesis (BBN) \cite{Seto2022,Takahashi2022bbn}, Tip of the Red Giant Branch method (TRGB) \cite{2019ApJ...882...34F,Li2022TRGB,Dhawan2022}, Mira variables \cite{2020ApJ...889....5H}, Tully-Fisher data \cite{Alestas2021b}, Megamaser \cite{2020ApJ...891L...1P}, Old Astrophysical Objects (OAO) \cite{2021NatAs...5..262J,2016ApJ...819..129O,Wei2022}, SNe Type II \cite{deJaeger2022}, gravitational signals and dark sirens \cite{	2021arXiv210402728G,2020ApJ...905...54G,1996gr.qc.....6079D,Li2021,Mozzon2021,Yang2021spaceborne,Fang2021,abbott2021gwskynetmulti,Gray2021,theligoscientificcollaboration2021constraints,Andreoni2021,Palmese2021,Zhu2021,Trott2021,Pol2022,Liu2022,Mukherjee2022,Ghosh2022,Jin2022,Huang2022gw,Pol2022sbgw,Liu2022darksirens,Chen2022,Wang2022,Califano2022,TorresOrjuela2022,Gupta2022greysirens}. Many papers consider the combination of different astrophysical probes for testing the cosmological models and investigating the Hubble tension \cite{	2016JCAP...10..019B,2019PhRvD.100j3501D,2018JCAP...04..051G,2019JCAP...04..036H,2021AJ....161..151K,2019ApJ...886L..23L,2020ApJ...895L..29L,2014ApJ...783..126P,2020arXiv200203599S,2017A&A...602A..73T,2021arXiv210510421V,2018PhRvD..97l3507W,2021arXiv210602963D,2019IJGMM..1650177H,2019A&A...628L...4L,2021JCAP...01..006D,2020A&A...643A..93H,2020SCPMA..6390402Z,2021MNRAS.504..300C,2021MNRAS.500.2227J,Mehrabi2021b,Jiang2021,Freedman2021,Galli2021,Fanizza2021,Arjona2021,Anderson2021,DiValentinodissecting,Wu2022,Umeh2022,Bora2022,Philcox2022,Shamir2022,Cao2022probes,Pal2022,Bulla2022,Lee2022,Cruz2022,Secco2022,Perivolaropoulos2022SH0ES,Bucko2022,Thakur2023}. Together with the assessed cosmological probes, new promising standardizable candles like the GRBs \cite{2019MNRAS.486L..46A,2013MNRAS.436...82D,2013ApJ...774..157D,2016ApJ...825L..20D,2017NewAR..77...23D,2017ApJ...848...88D,2007RSPTA.365.1363L,2000ApJ...543..722L,2015ApJ...806...44P,2018AdSpR..62..662S,2020A&A...641A.174L,2021ApJ...907..121T,2021MNRAS.501.3515M,2014ApJ...783..126P,Dainotti2008,Duncan2001,DallOsso2011,Cardone2009,Cardone2010,Capozziello2011,Dainotti2010,Dainotti2011a,Dainotti2013a,Dainotti2015a,Dainotti2015b,DelVecchio2016,DainottiDelVecchio2017,Dainotti2016,Dainotti2017a,Dainotti2017b,Dainotti2020a,Dainotti2020b,DainottiAmati2018,Dainotti2021d,Dainotti2018,Cucchiara2011,Rowlinson2014,Rea2015,Stratta2018,DainottiPetrosianBowden2021,Yonetoku2004,Ghirlanda2010,Amati2002,Atteia1997,Atteia1998,LiangZhang2005,Dainotti2021proceeding,Ito2019,DainottiDelina2021,Cao2022Platinum,wang2021prospects,Luongo2022grb,Liu2022grb,Lenart2021Gammaproceeding,Srinivasaragavan2020}, the QSOs \cite{	2019NatAs...3..272R,2011ApJ...743..104S,2013ApJ...764...43S,2021arXiv210503965Z,Wang2021,bargiacchi2021quasar,Gupta2022,Li2022qso,Simon2022eBOSS,Huang2022lensedqso}, the active galactic nuclei (AGNs) \cite{Ingram2021,Narendra2021,Lu2021}, the planetary nebulae luminosity function (PNLF) \cite{2021ApJ...916...21R}, the Fast Radio Bursts (FRBs) \cite{Wu2021,Boone2022,Liu2022FRB}, J-Branch Asymptotic Giant Branch stars (JAGB) \cite{Lee2021JAGB}, and the extragalactic radio jets \cite{Hsiao2022} are promising tools for casting more light on the open cosmological problems.\\
Other suggestions in literature are given by the model-independent approaches \cite{2020arXiv201110559R,2018PhRvD..98h3526S,2019MNRAS.485.2783L,2019JCAP...07..005Z,2020CQGra..37w5001C,2021JCAP...03..034K,2020ApJ...897..127W,2021EPJC...81...36M,2021MNRAS.501.5714W,sun2021influence,renzi2020look,Reyes2021,Mehrabi2021a,Bernardo2022nonpar,Dialektopoulos2021,Renzi2022DDR,Hu2022,Yang2022,Lemos2022}, cosmography \cite{2020PhRvR...2a3028C,2020PhRvD.102l3532Y,2020MNRAS.491.4960L,2020ApJ...900...70R,Desouza2021,Mehdi2021,Giada2021,Shajib2022TDCOSMO,Liu2022cosmography,Seymour2022,Treu2022,Pourojaghi2022,Birrer2022}, cosmic triangles schemes \cite{2021PhRvD.103j3533B}, machine learning techniques \cite{Narendra2021,Bengaly2022}, the ABC algorithm \cite{Bernardo2022}, the alternative resolution of Friedmann equations \cite{SandovalOrozco2022,Jusufi2023} and the discussion of distance measurement in cosmology \cite{Greene2021}.\\
Finally, it is essential to discuss also the presence of biases in the collected data samples \cite{1920MeLuF..96....1M,EfronPetrosian1992,Parnovsky2021a,Baldwin2021,Fleury_2017,Adamek2019,Blanchard2022} since, if this is the case, the biased variables may lead to unreliable cosmological results.\\
A more complete description of the Hubble constant tension approaches and the basics of cosmological models can be found in the review papers and books \cite{	2001LRR.....4....1C,2021arXiv210301183D,2008Natur.452..158E,1990eaun.book.....K,2003PhRvL..90i1301L,2011prco.book.....M,1971phco.book.....P,2020NatRP...2...10R,2020PhRvD.102b3518V,2019NatAs...3..891V,2008cosm.book.....W,1935ApJ....82..284R,2021arXiv210505208P,1961PhRv..124..925B,jordan1955schwerkraft,Cai2021,paul2021,Moresco2022,CyrRacine2021review,Schoneberg2021,Perivolaropoulos2021review,DivalentinointertwinedI,DivalentinointertwinedII,DivalentinointertwinedIII,DivalentinointertwinedIV,Aluri2022review}.\\
In the current work, we will describe the analysis performed through the use of SNe Ia \cite{2021ApJ...912..150D} and SNe Ia combined with Baryon Acoustic Oscillations (BAOs) \cite{Dainotti2022SNe} to investigate the Hubble constant tension at local redshifts ($z$), given that the redshift of SNe Ia is $z_{SNeIa}\leq2.26$. We will then show how the GRBs and QSOs can contribute to the estimation of cosmological parameters and discuss their application in the $H_0$ tension problem.

\section{Data analysis}
\subsection{The Hubble tension in the Pantheon sample: part I}
\begin{figure}[h!]
    \centering
    \includegraphics[scale=0.2]{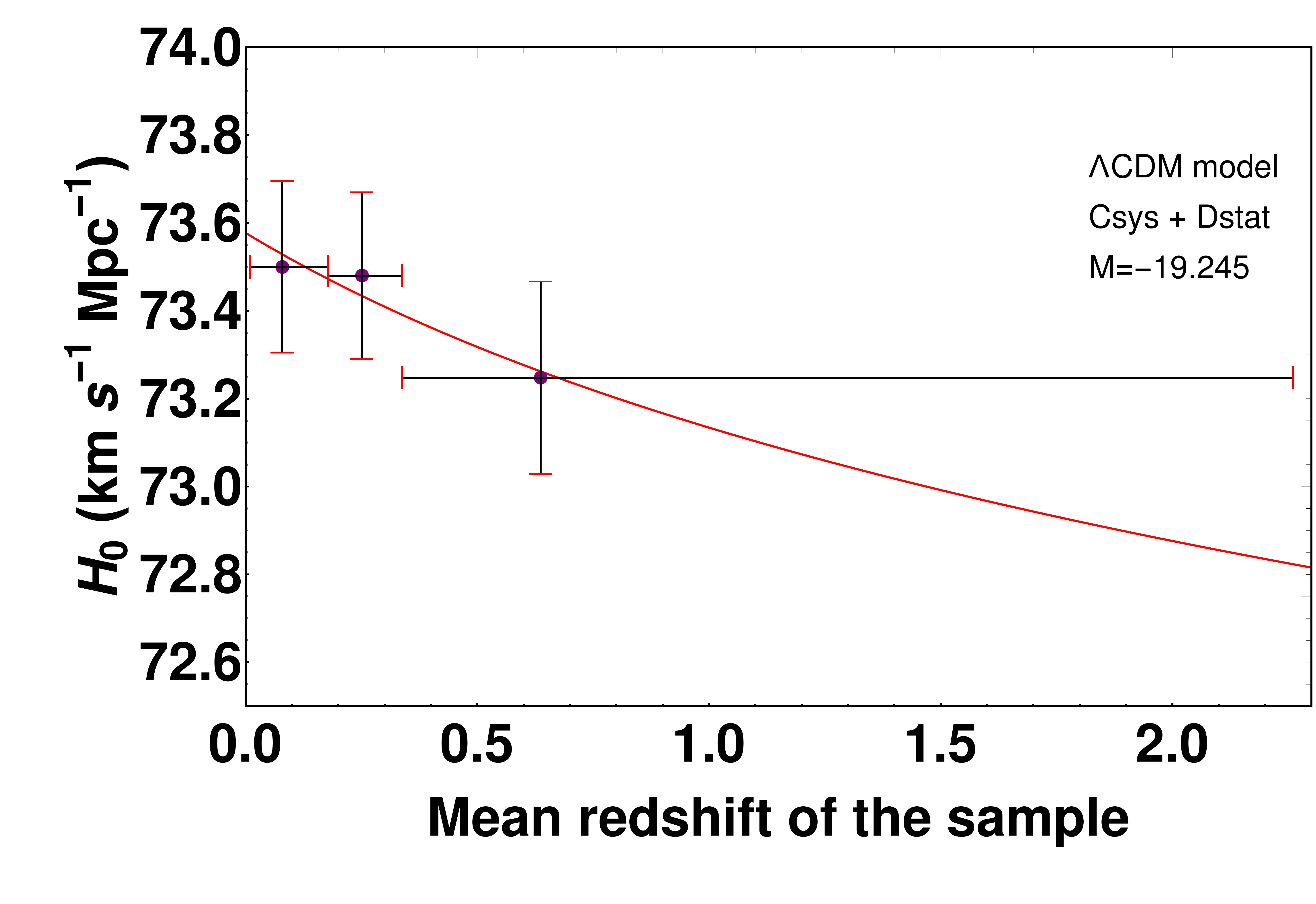}
    \includegraphics[scale=0.2]{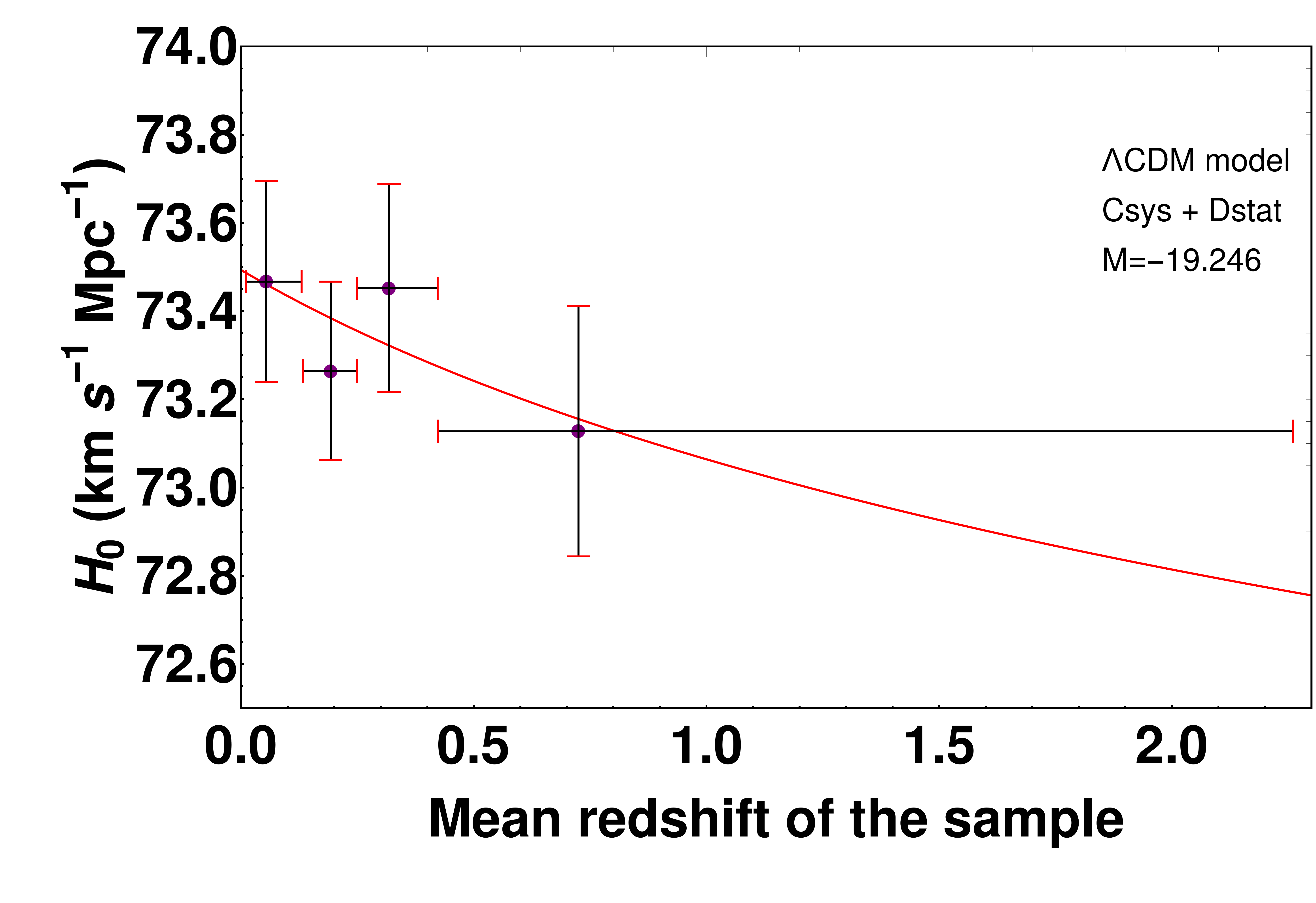}
    \caption{\textbf{Upper panel.} The decreasing trend of $H_0$ in the 3 bins division of the Pantheon sample. \textbf{Lower panel.} The decreasing trend of $H_0$ in the 4 bins division of the Pantheon sample. Both the plots are taken from \cite{2021ApJ...912..150D}.}
    \label{fig1}
\end{figure}

In this subsection, we summarize the results obtained in \cite{2021ApJ...912..150D}.\\
To investigate the $H_0$ tension through SNe Ia, a \textit{binning approach} was adopted. The Pantheon sample \cite{2018ApJ...859..101S}, a collection of 1048 SNe Ia with a redshift range in $0<z_{SNeIa}\leq2.26$, was divided into 3 and 4 equally populated SNe bins ordered in redshift. The division into 3 bins is the best option for giving each bin a reliable statistical weight, while the 4 bins have been extracted to compare our results with the ones in \cite{2020PhRvD.102b3520K}. The value of SNe Ia fiducial absolute magnitude $M$ was assumed such that the local $H_0$ in the first bin of each division (first out of 3 and first out of 4) is calibrated on the value of $H_0=73.5\,km\,s^{-1}\,Mpc^{-1}$. \\ 
For each bin, a 1-dimensional Monte Carlo Markov Chain (MCMC) analysis has been performed using the \textit{Cobaya package}\footnote{\url{https://cobaya.readthedocs.io/en/latest/index.html}} \cite{Torrado:2020dgo}, allowing only $H_0$ to vary in both the $\Lambda$CDM and $w_{0}w_{a}$CDM \cite{2001IJMPD..10..213C} models. Subsequently, the $H_0$ values obtained have been fitted with the following parametric form:

\begin{equation}
f(z)=\frac{\tilde{H_0}}{(1+z)^\alpha},
\label{eq1}
\end{equation}

where $\alpha$ is the evolutionary parameter for $H_0$ with redshift and $\tilde{H_0}$ is the local value of the Hubble constant obtained with the fitting, namely $f(z=0)=H_0(z=0)$.\\
The results show that the $H_0$ has a mild decreasing trend with redshift, with an $\alpha$ parameter in the order of $10^{-2}$. More specifically, from the test of the $\Lambda$CDM cosmological model, we found out that $\alpha=0.009\pm0.004$ in the 3 bins division, thus $\alpha$ is compatible with zero only in 2.0 $\sigma$. In 4 bins, the parameter $\alpha=0.008\pm0.006$ is compatible with zero in 1.5 $\sigma$. The results of $w_{0}w_{a}$CDM model are compatible in 1 $\sigma$ with the ones of $\Lambda$CDM model and will not be reported. The fitting for 3 and 4 bins in the $\Lambda$CDM model are visible in the upper and lower panels of Figure \ref{fig1}, respectively.

\subsection{The Hubble tension in the Pantheon sample: part II}
In this second subsection, the approach adopted in \cite{Dainotti2022SNe} is reported.\\
To investigate if this trend is still visible when leaving more cosmological parameters free to vary and adding another cosmological probe, the BAOs, we performed the following modifications to the approach shown in \cite{2021ApJ...912..150D}:
\begin{itemize}
    \item the $H_0$ and $\Omega_{0m}$ were both allowed to vary in the $\Lambda$CDM model;
    \item the same consideration has been done for $H_0$ and the $w_{a}$ parameters in the $w_{0}w_{a}$CDM model, being $w(z)=w_{0}+w_{a}*z/(1+z)$ the equation of state in the CPL parametrization;
    \item A further cosmological probe has been added; namely, the BAOs from \cite{Sharov_2018};
    \item The Pantheon sample combined with BAOs was divided into 3 bins to avoid the statistical fluctuations dominating the analysis;
    \item The fiducial value of SNe Ia absolute magnitude is $M=-19.25$, like in the original release of Pantheon \cite{2018ApJ...859..101S} where the value of $H_0=70\,km\,s^{-1}\,Mpc^{-1}$.
\end{itemize}

\begin{figure}[h!]
    \centering
    \includegraphics[scale=0.55]{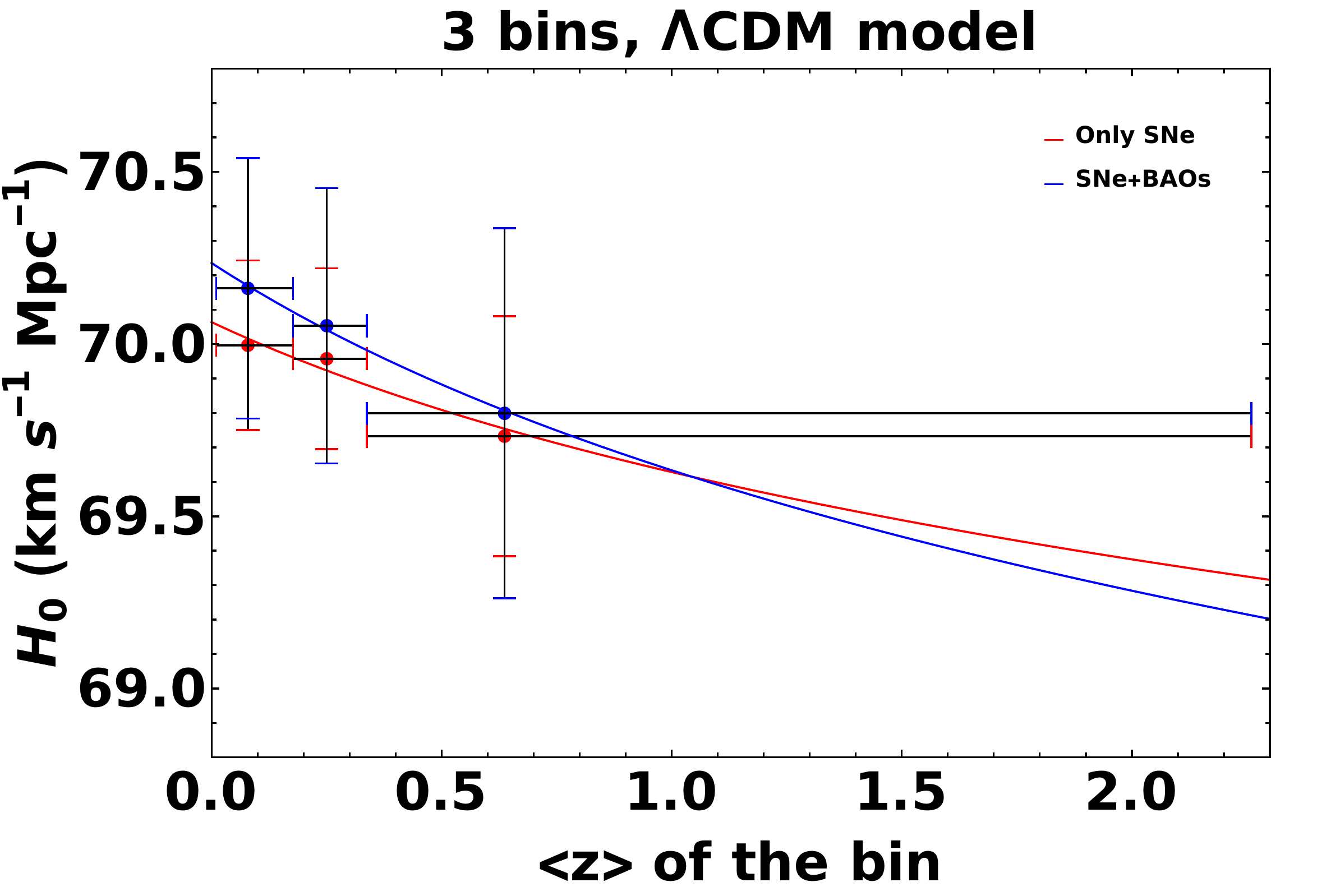}
    \includegraphics[scale=0.5]{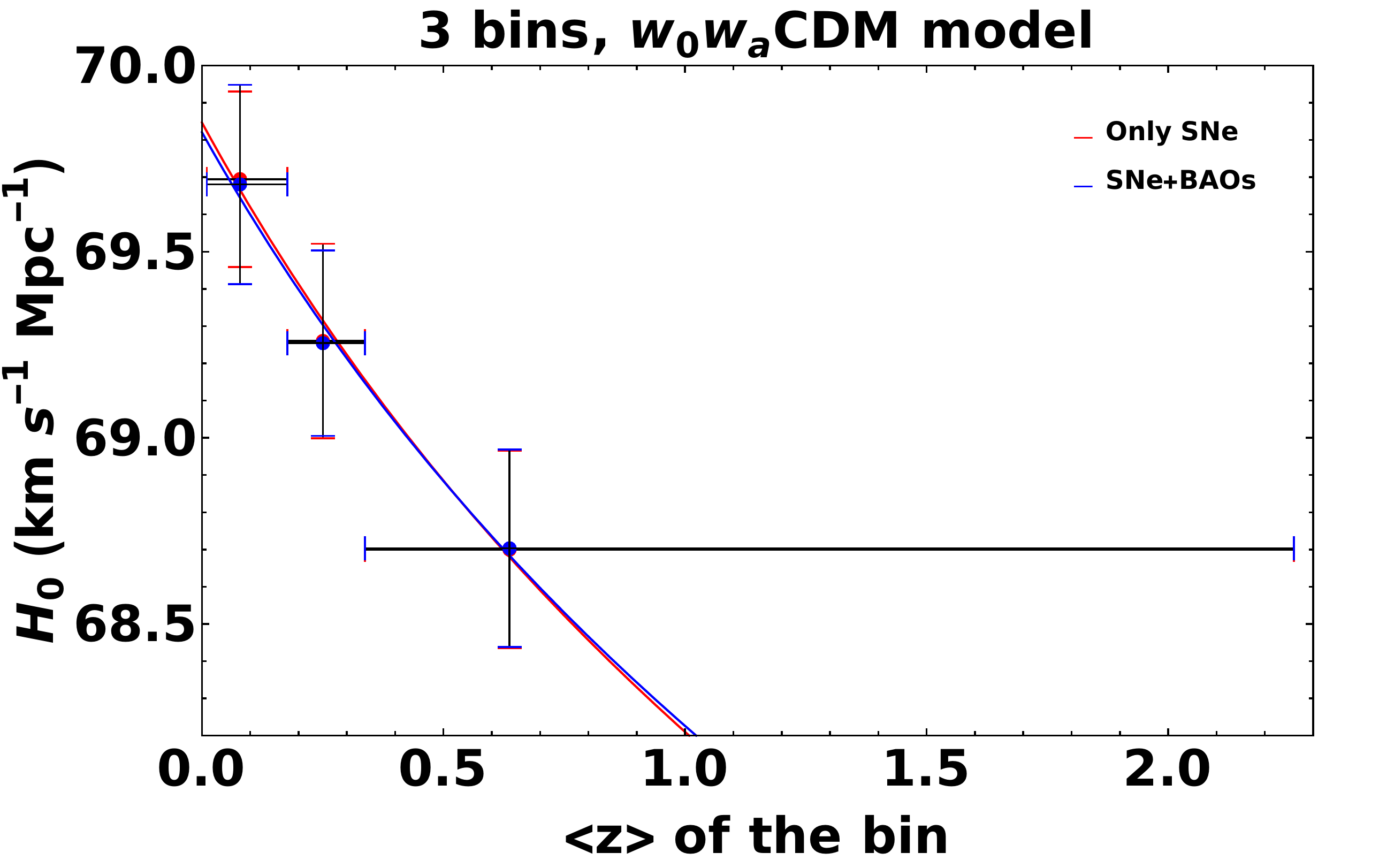}
    \caption{\textbf{Upper panel.} The decreasing trend of $H_0$ in the 3 bins division of the Pantheon sample, adding the BAOs and also allowing $\Omega_{0m}$ to vary. In red is the case where only SNe Ia are present in the bins, while in blue is the case of SNe Ia + BAOs present. \textbf{Lower panel.} The decreasing trend of $H_0$ in the 4 bins division of the Pantheon sample, with BAOs and the variation of $H_0$ and $w_a$. The same color coding of the upper panel is adopted here. The plots are extracted from \cite{Dainotti2022SNe}.}
    \label{fig2}
\end{figure}

Following the aforementioned approach, the results still show a decreasing trend for $H_0$ with redshift and an $\alpha$ in the order of $10^{-2}$. According to the $\Lambda$CDM model, the value of $\alpha=0.008\pm0.006$ is compatible with zero in 1.2 $\sigma$, while in the $w_{0}w_{a}$CDM model $\alpha=0.033\pm0.005$ is compatible with zero in only 5.8 $\sigma$. The fitting for the $\Lambda$CDM and the $w_{0}w_{a}$CDM models are reported in the upper and lower panel of Figure \ref{fig2}, respectively.\\

\section{Towards a physical interpretation of the Hubble constant trend}
We now discuss the possibility of a physical interpretation of the observed profile of $H_{0}\left(z\right)$. In other words, we investigate new physics that could account for a redshift evolution of the $\Lambda$CDM model, resulting in a rescaling of the luminosity distance, which reproduces an effective evolution of the Hubble constant, i.e., $H_0(z) \sim (1+z)^{-\alpha}$. 

\begin{figure}[h!]
    \centering
    \includegraphics[scale=0.35]{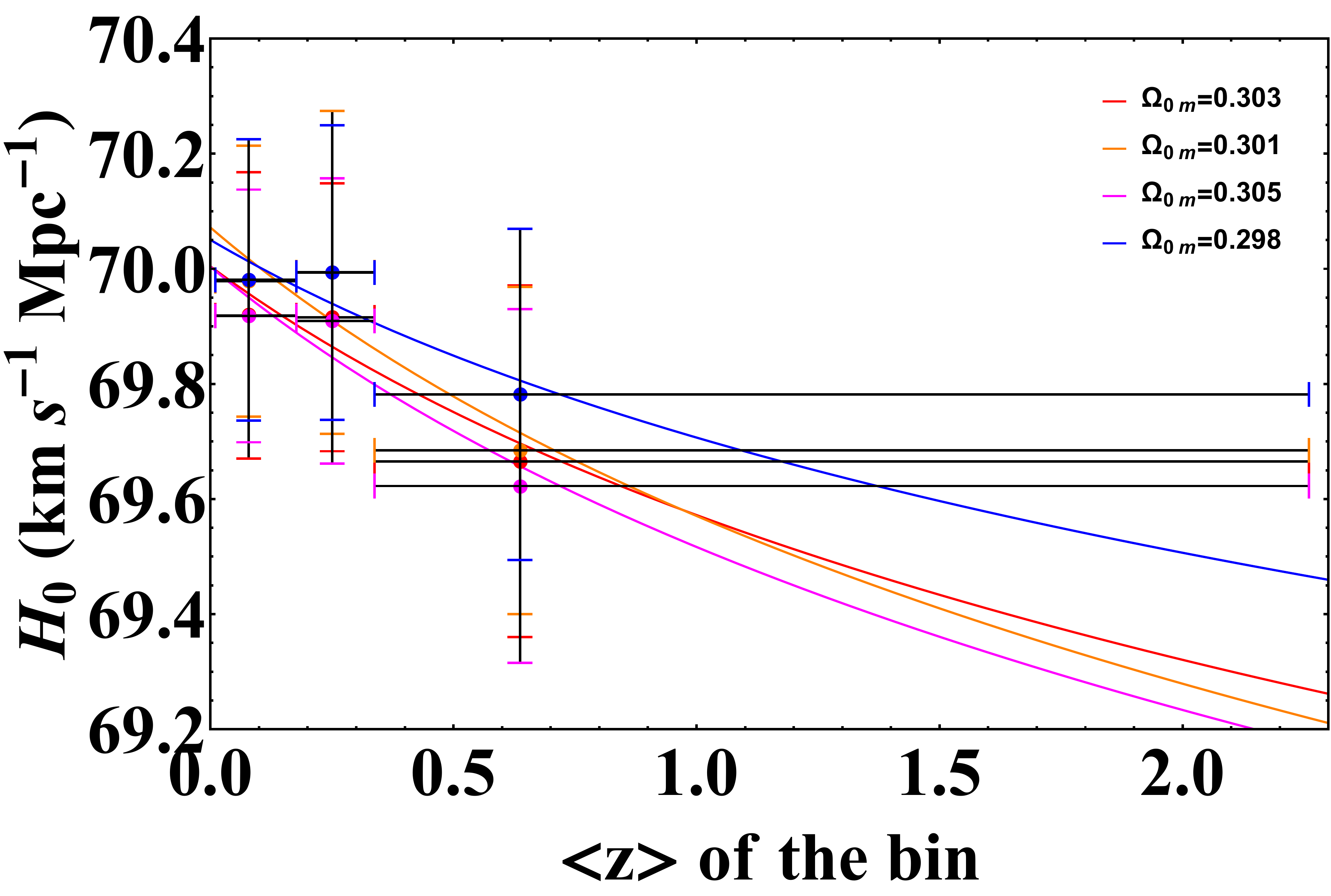}
    \caption{$H_0$ versus $z$ plot in the case of $|F_{R0}|\leq10^{-7}$, according to four different values of $\Omega_{0m}=0.298,0.301,0.303,0.305$ and combining SNe Ia+BAOs (plot taken from \cite{Dainotti2022SNe}).}
    \label{fig3}
\end{figure}

In \cite{2021ApJ...912..150D}, it was argued that a natural framework to obtain the desired effect is the $f(R)$ modified gravity in the Jordan frame \cite{Nojiri-Odintsov2007,Nojiri:2010wj-unified,Sotiriou-Faraoni2010,Faraoni:2010pgm,Capozziello:2011et,Nojiri:2017ncd-nutshell}, simply because it provides a rescaling of the Einstein constant, precisely what it would need for an additional redshift dependence for the $\Lambda$CDM cosmological dynamics. The quantity that could be responsible for this rescaling of the Einstein constant is the non-minimally coupled scalar field $\phi$ to standard gravity, as it emerges in the Jordan frame. The dynamics of $\phi$ are influenced by a potential term $V\left(\phi\right)$, which is entirely fixed by the particular form of the $f(R)$ function we are considering. However, this perspective was investigated in detail in \cite{Dainotti2022SNe}, where some discrepancies for the observed $H_0(z)$. On the one hand, the use of the Hu-Sawicki model \cite{Hu:2007nk,2009PhRvD..79l3516M}, namely one of the most reliable $f(R)$ proposal for the description of a dark energy dynamics, was unable to reproduce the amended luminosity distance for removing the dependence $H_0(z)$ from data, see the plot in Figure \ref{fig3}; on the other hand, the possibility for slow-rolling dynamics of the scalar field encountered some difficulties. Only in the dark energy-dominated era a viable model has been derived. 

Here, we briefly elucidate the theoretical paradigm that in \cite{Schiavone:2022wvq} led to propose a revised $\Lambda$CDM dynamics from $f(R)$ gravity, which reproduces the desired $H_0(z)$ behavior. 
The starting point is the modified Friedmann equation for an isotropic late Universe, as it emerges in the Jordan frame, i.e.: 

\begin{equation}
	H^2 = \frac{\chi\rho_m}{3\phi} - H\frac{\dot{\phi}}{\phi} + \frac{V}{6\phi}
	\, ,
	\label{gtm1}
\end{equation}

where $\rho_m$ denotes the energy density of the matter component in the late Universe, setting $c=1$.

Now, without loss of generality, we can express the scalar field as $\phi = \phi (z)$, and we can assume that for a range $0 < z \ll 1$, the potential term remains sufficiently flat to be approximated with a constant value $V\simeq 2\chi \rho_{\Lambda}\simeq \textrm{const}$ with $\rho_{\Lambda}$ being the value of the late universe vacuum energy. Hence, it is immediate to recognize that the Friedmann equation \eqref{gtm1} can be rewritten as:

\begin{equation}
	H^2 = \frac{1}{\phi - (1+z) \frac{d\phi}{dz}} \frac{\chi}{3} \left(\rho_m + 
	\rho_{\Lambda}\right) 
	\, .
	\label{iom1}
\end{equation}

We see that the desired rescaling of the Einstein constant is ensured by the factor $1/[\phi - (1+z)d\phi /dz]$, which must equate the observed profile $1/(1+z)^{2\alpha}$. This request fixes $\phi (z)$. Furthermore, the necessary consistency of the dynamics equation governing $\phi$ is checked by evaluating the derivative of the scalar field potential $V\left(\phi\right)$ for low redshifts. 

In \cite{Schiavone:2022wvq}, it was shown that the proposed scenario consistently works, reproducing both an effective evolution of the Hubble constant and a current cosmic accelerated phase. Furthermore, the same picture was implemented on a purely numerical level without imposing the existence of a flat region of the potential, which naturally emerges for $z<0.3$, i.e., in the dark energy-dominated Universe.

\section{GRBs and QSOs: promising cosmological probes}
\begin{figure}
    \centering
    \includegraphics[scale=0.32]{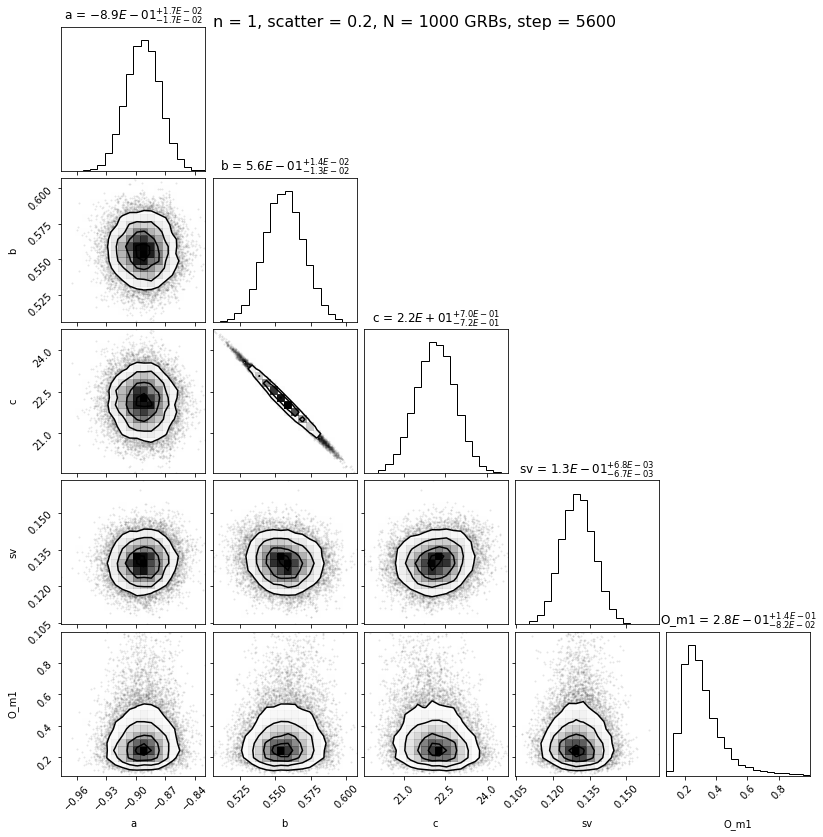}
    \includegraphics[scale=0.32]{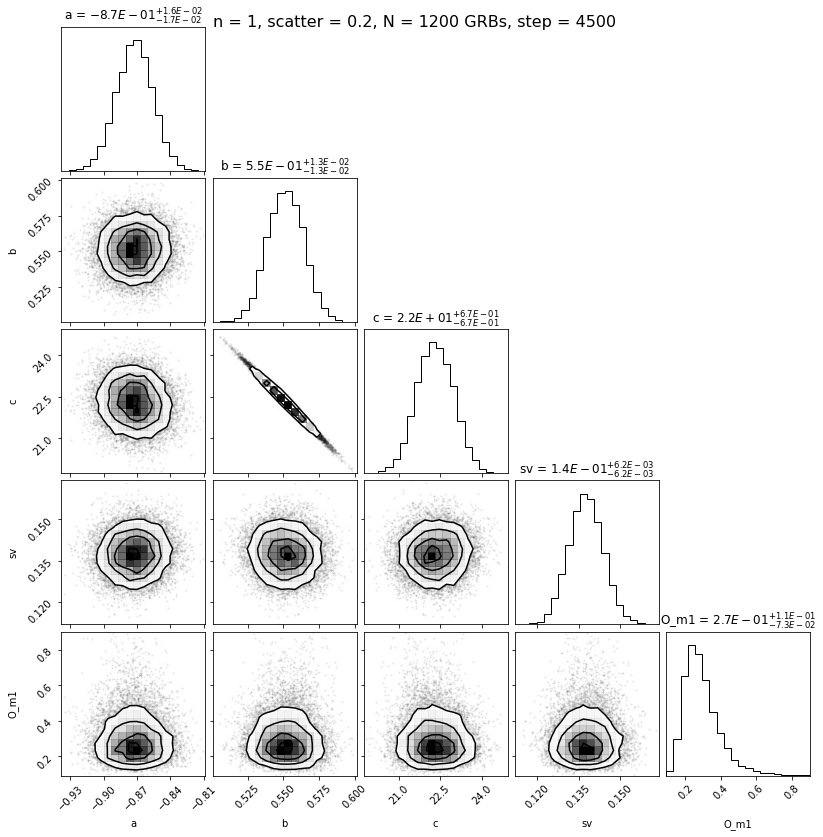}
    \caption{\textbf{Upper panel.} The simulation of 1000 GRBs according to the fundamental plane properties, with 5600 steps elapsed. \textbf{Lower panel.} The simulation of 1200 GRBs, with 4500 steps elapsed. The plots are taken from \cite{DainottiNielson2022}.}
    \label{fig4}
\end{figure}

GRBs are the most energetic phenomena observed in the universe after the Big Bang. Their origin is cosmological, and they arise from the collapse of massive stars in the case of Long GRBs (with intrinsic duration $>2\,s$), while in the case of Short GRBs ($<2\,s$) the progenitor is believed to be the merging of two compact objects - such as two neutron stars or a neutron star with a black hole. GRBs are observed up to high redshifts; the distance record is currently held by GRB 090429B with a redshift value $z=9.4$ \cite{Cucchiara2011}. It is necessary to invoke reliable correlations between the luminosity or energy of the GRBs and other physical parameters independent of the cosmological model to use GRBs as standardizable candles. \\
In the past years, many contributions in this field have been provided focusing on the prompt properties of GRBs \cite{Amati2002,Yonetoku2004,Ghirlanda2010}. More recently, the idea of considering the afterglow properties for the search of tight and unbiased correlations has led to the discovery of the 2-dimensional Dainotti relations \cite{Dainotti2008,Dainotti2010,Dainotti2011a,Dainotti2013a,Dainotti2015a,Dainotti2015b,DelVecchio2016,Dainotti2020b,LevineDainotti2022} between the end of the GRB plateau emission $T^{*}_a$ and the luminosity at the end of the plateau $L_a$ in the X-rays, optical, and radio wavelengths. Adding the peak prompt luminosity, $L_p$, the so-called \textit{fundamental plane relation} or (Dainotti relation) is unveiled \cite{2013MNRAS.436...82D,2014ApJ...783..126P,Dainotti2016,Dainotti2017a,Dainotti2017b,2017NewAR..77...23D,Dainotti2018,DainottiAmati2018,Dainotti2020a,Dainotti2021closure,Dainotti2021Fermi,Dainotti2021AASproceeding,DainottiLenart2023,Dainotti2022binned,Cardone2009,Cardone2010,Cao2022,Cao2022Platinum,Lenart2021Gammaproceeding,Srinivasaragavan2020}. This correlation can be expressed through the following equation:

\begin{equation}
    log_{10}L_a=a*log_{10}T^{*}_a+b*log_{10}L_{p}+c,
    \label{eq4}
\end{equation}

being $a,b$ the parameters of the plane and $c$ the renormalization constant. It is important to stress that the physical observables $T^{*}_a$, $L_a$, and $L_p$ have been corrected for selection biases and redshift evolution effects through the Efron and Petrosian method \cite{EfronPetrosian1992}. First, for the fundamental plane reference, the \textit{platinum sample} of X-ray GRBs is chosen \cite{Dainotti2020a}. Then, in \cite{DainottiNielson2022} the simulations of 1000 GRBs and 1200 GRBs according to the platinum sample properties have been performed. For each of the two simulated samples, a MCMC analysis with the D'Agostini method \cite{1995NIMPA.362..487D} has been done varying the parameters $a,b,c,\sigma,\Omega_{0m}$, where $\sigma$ is the intrinsic scatter of the fundamental plane relation in Equation \ref{eq4}. The results show how in the case of 1000 GRBs, a value of $\Omega_{0m}=0.280\pm0.111$ is obtained, while in the 1200 GRBs simulated sample, the same parameter is $\Omega_{0m}=0.270\pm0.092$. The results are reported in Figure \ref{fig4}. According to the rate of GRBs with plateau observed by the Swift satellite, it is expected that by 2030 the observations of GRBs with properties similar to the ones from the platinum sample will lead to a standalone estimation of the $\Omega_{0m}$ parameters comparable with the one reported in \cite{Conley2011}.

QSOs are very bright active galactic nuclei observed up to redshift $z=7.642$ \cite{Wang2021}. Similarly to GRBs, to use QSOs as distance measurement tools a statistically reliable correlation between their luminosity and other parameters is needed. To this end, the use of Risaliti \& Lusso (RL) relation \cite{Lusso2016} between the X-ray and ultraviolet fluxes of quasars represents a reliable candidate. The RL relation can be expressed with the following: 

\begin{equation}
    log_{10}L_X=g*log_{10}L_{UV}+b,
\end{equation}

denoting with $sv$ the intrinsic scatter of the correlation. The unbiased nature of this correlation for QSO has been proved in \cite{DainottiBargiacchi2022,LenartBargiacchi2022}. The RL relation has been applied to QSOs \cite{LenartBargiacchi2022}: assuming a flat $\Lambda$CDM model, combining QSOs and SNe Ia, the results yield different values of $H_0$ which are in middle way between the results of SNe Ia and Planck data.


This outcome shows how the use of QSOs may significantly impact the study of the Hubble tension.

\section{Conclusions}
The $H_0$ tension is a long-standing problem that still challenges our current knowledge of astrophysics and cosmology. In particular, the approach adopted for the analysis of the SNe Ia Pantheon sample shows a mild decreasing trend of $H_0$ with redshift. This trend is present in the data, regardless of the number of bins in which the Pantheon sample has been divided or the number of parameters allowed to vary or, even more, the presence of a further cosmological probe (BAO) even binned.\\
These outcomes could be due to the presence of hidden astrophysical biases or redshift evolution effects in the SNe Ia data. One of the possible causes can be the evolution of the SNe Ia stretch parameter with redshift which has already been discussed in \cite{2021A&A...649A..74N}.\\ 
In case results are not due to biases or astrophysical effects, they can be explained through alternative theoretical scenarios, e.g., the modified gravity theories like the $f(R)$ theories of gravity. Since the Hu-Sawicki model has not fully accounted for the observed $H_0$ trend, it is likely that this effect could be still due to modified gravity dynamics, but with different features from the Hu-Sawicki model.\\
Finally, we have also shown that GRBs and QSOs are promising tools for the measurement of cosmological distances and the estimation of cosmological parameters.
In the case of GRBs, the analysis of 1200 events simulated from the fundamental plane relation gives a value $\Omega_{0m}=0.270\pm0.092$, confirming the possibility for the GRBs to reach, by 2030, the same precision on this parameter as the estimation in \cite{Conley2011}.\\
For the QSOs, when these are combined with SNe Ia, mixed outcomes about the Hubble constant are found.\\
These phenomena extend the Hubble diagram up to $z=7.642$ and $z=9.4$ in the cases of QSOs and GRBs, respectively: these values are far beyond the current observational limit of SNe Ia, being the furthest SNe Ia observed at $z=2.26$. This important feature of GRBs and QSOs suggests how these probes can be used to cast more light on open cosmological problems, in particular, the Hubble tension.



\bibliographystyle{JHEP}
\bibliography{H0_papers}

\end{document}